\documentclass[aip,reprint,twocolumn,author-year]{revtex4-1}

\usepackage{amsmath}
\usepackage{amssymb}
\usepackage{amsfonts}
\usepackage{dsfont}
\usepackage{graphicx}
\usepackage{color}
\usepackage{subfigure}
\usepackage[latin2]{inputenc}
\usepackage{hyperref}
\usepackage{multirow}

\usepackage{scalefnt}
\usepackage{placeins}

\usepackage[normalem]{ulem} 
\renewcommand{\vec}{\bm}

\begin{document}

\title{\LARGE Estimating correlation and covariance matrices by weighting of market similarity}
\author{Michael C. M\"unnix}
\email{michael@muennix.com}
\author{Rudi Sch\"afer}
\affiliation{Department of Physics, University of Duisburg-Essen, Duisburg, Germany}
\author{Oliver Grothe}
\affiliation{Department of Economic and Social Statistics, University of Cologne, Germany}

\begin{abstract}
We discuss a weighted estimation of correlation and covariance matrices from historical financial data. To this end, we introduce a weighting scheme that accounts for similarity of previous market conditions to the present one. The resulting estimators are less biased and show lower variance than either unweighted or exponentially weighted estimators.

The weighting scheme is based on a similarity measure which compares the current correlation structure of the market to the structures at past times. Similarity is then measured by the matrix 2-norm of the difference of probe correlation matrices estimated for two different times.
The method is validated in a simulation study and tested empirically in the context of mean-variance portfolio optimization. In the latter case we find an enhanced realized portfolio return as well as a reduced portfolio volatility compared to alternative approaches based on different strategies and estimators.

\end{abstract}


\maketitle

\section{Introduction}
Good estimates of the correlation and covariance matrices of financial returns are central for a wide range of applications such as risk management, option pricing, hedging and capital allocation. For example in risk management applications, they directly affect the calculation of the value at risk or the expected shortfall. In the context of capital allocation, the correlation structure is key in the classical portfolio optimization problem, as shown in the seminal work of \cite{markowitz52}.

Generally, the quality of the estimated matrices increases with the length of the time series, i.e., the amount of data used. For small datasets the matrices have a large variance and may even be singular or indefinite. In financial context, however, using long time series results in biased estimates of the correlation structure, since the dependence of asset returns is not constant in time (see, e.g., \cite{king90} for an early review).

The problem is that standard estimators equally weight all parts of the dataset. By consequence, out-of-date and improper information highly affect the estimates.
This paper tackles this problem by introducing a new weighted estimator of the correlation or covariance matrix. This estimator makes use of enough data to adequately limit its variance but - in order to minimize its bias - focuses only on parts of the data where the market is in similar market conditions, i.e., it exhibits the same correlation structure.

To reduce the effects of time changing structures, common approaches in the literature choose time intervals where the structures are approximately constant. Examples of such approaches are exponentially weighted estimators like the RiskMetrics estimators (see, e.g., \cite{RiskMetrics})
or the estimators discussed in \cite{lee-s03}.
Since these estimators only use a small part of the data, they show a large variance. Moreover, whenever the number of effectively used observations is not large compared to the dimensions of the time series, estimated correlation and covariance matrices may be regarded as completely random. \cite{laloux99} showed in an empirical example that in such cases $94\%$ of the spectrum of estimated correlation matrices equal the spectrum of random matrices and only their largest eigenvalues may be estimated adequately.

Solutions to this problem involve reducing the dimensionality of the problems by imposing some structure on the correlations, e.g., by using factor models or shrinkage estimators as in \cite{ledoit04} or by noise reduction techniques, e.g., Random Matrix Filtering (see \cite{plerou03}) or Power Mapping (\cite{schaefer08}). Other approaches reduce the dimensionality by using conditional models of the correlation matrices going back to the work of \cite{bollerslev86}.
A short overview of these practices may be found in \cite{andersen07}.

With the availability of intraday high frequency financial data, it was expected that finer sampled data would effectively enlarge the datasets and improve estimates of parameters. However, when return data is observed on shorter time intervals, it is contaminated by market microstructure effects. These effects influence estimators and induce bias and noise (see, e.g., for a recent discussion \cite{bandi08}). Possible reasons include asynchrony and decimalization effects (see, e.g., \cite{muennix09b,muennix10a}).

Since the amount of data for the estimation may only be increased by either considering a longer time period or by sampling on higher frequencies, the mentioned properties of financial time series limit the amount of usable data.
Longer time intervals bias the estimators due to the time changing nature of the matrices. Higher frequencies intensify the effects of the market microstructure on the estimators.

In this paper, we circumvent these limits. We propose to enlarge the amount of usable data by adaptively including different parts of the time series with similar correlation structures into the estimator. We therefore introduce a similarity measure which measures the degree of similarity between days of the time series based on probe correlation estimations.
We demonstrate the application of the measure on assessing similarities on stock returns from the S\&P 500 index. The measure reliably detects regime changes in the data as well as the special market situation during the financial crisis in $2008$.

The similarity measure enables us to construct a weighting scheme for correlation or covariance estimators that attaches high weights on similar parts of the data and suppresses distortions. In a simulation study, we demonstrate that these similarity weighted estimators show smaller bias and variance than unweighted or exponentially weighted estimators. The results hold for constant as well as for dynamic correlation structures in the data.
In a real data application we apply our estimator to covariance estimation in the context of mean-variance portfolio optimization. We use time-series of stocks from the S\&P 500 index and randomly choose stocks to build up portfolios. We show that optimal portfolios which are based on the similarity weighted covariance estimator outperform alternative approaches with respect to realized volatility and realized return.

The paper is organized as follows. Section \ref{sec.similarity} introduces the measure of similarity. In section \ref{sec.weightcorr}, a similarity based weighting scheme for estimators of correlation or covariance is constructed. Section \ref{sec.simStud} contains a simulation study analyzing variance and bias of the resulting estimators. In section \ref{sec.emp} we empirically apply the estimators in the context of mean-variance portfolio optimization. Section \ref{sec.concl} concludes.

\section{Measuring Market Similarity}
\label{sec.similarity}
\begin{figure}[tp]
	\centering
		\includegraphics[width=0.48\textwidth]{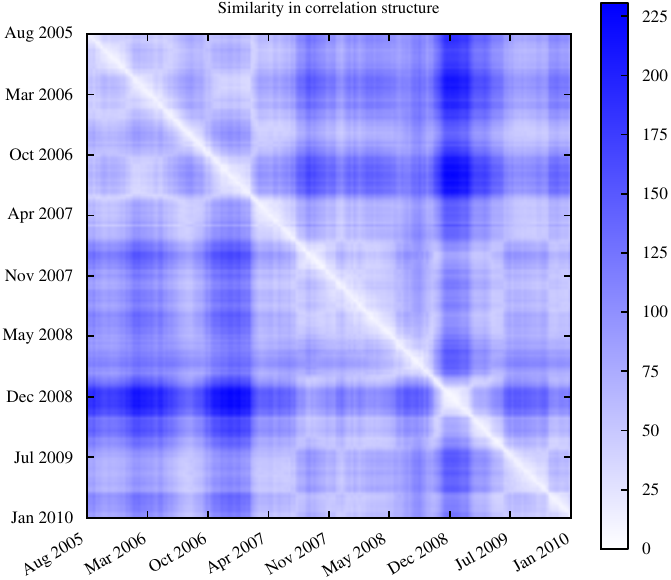}
	\caption{Illustration of of the $\zeta^{50}$ similarity measure for the correlation structure of the S\&P 500 index from 2005 to the beginning of 2010. Each point of the graphic reflects the degree of similarity between the days at its coordinates. The dark shaded areas indicate a correlation structure that is not similar to any other period before or after, while the white areas indicate high values of similarity. The region past Oct $2008$ can clearly be identified as the beginning of the financial crisis in $2008$. Furthermore, in Feb $2007$ the correlation structure of the assets changes.}
	\label{fig:simmatrix}
\end{figure}
We measure the degree of similarity $\zeta$ in the market's correlation structure by the norm of the difference of the correlation matrices $\mathbf{C}(t_{1})$ and $\mathbf{C}(t_{2})$ of the times $t_1$ and $t_2,$ i.e.,
\begin{align}
\zeta(t_{1},t_{2}) = \big | \big|\mathbf{C}(t_{1})-\mathbf{C}(t_{2}) \big | \big|_{2}\ ,
\label{eq:simmeasure}
\end{align}
where $||\mathbf{C}||_{2}$ represents the induced matrix $2$-norm of the real valued matrix $\mathbf{C}$, which is the square root of largest eigenvalue of the matrix $\mathbf{C}'\mathbf{C}$.

The correlation matrices $\mathbf{C}(t_{1})$ and $\mathbf{C}(t_{2})$ are estimated on a backward-looking rolling window of length $L.$ The window length $L$ will be indicated by a superscript, i.e., $\zeta^{L}.$ If outliers are present in the data, the estimates are based on Spearman's rank correlation instead of Pearson's product moment correlation as this estimator is more robust to non-normal distributions.
Since the estimates should be unbiased for time varying correlations, the use of small window lengths is recommended.
As discussed in \cite{laloux99}, this results in noisy estimates of the matrices and only the largest eigenvalues of the matrices are adequately estimated. However, the similarity measure (\ref{eq:simmeasure}) is based on the $2$-norm and thus depends only on this largest eigenvalue, which can be estimated even for small values of $L.$

 Figure \ref{fig:simmatrix} illustrates the evolution of the similarity measure $\zeta^{50}$ for the example of the $471$ assets, that were continuously in the S\&P 500 index between Aug $2005$ and Jan $2010$.
The similarity measure is evaluated for every day between Aug $2005$ and Jan $2010$ and depicted as a matrix. The axes represent time, therefore the evolution of the market related to a specific point in time is given by the upright (or vertical) intersection through this point. Darker regions on this intersection are less similar and brighter regions more similar to the situation at the specific point in time. In this illustration, the the financial crisis causes a shaded area from Oct 2008 to Mar 2009.
The correlation structure in this period is completely different from any period before. After this period we find the market stabilizing: The correlation structure becomes similar to previous market states again. Beside the financial crisis, we can find regions for any point in which the correlation structure was similar and regions where it was different.

\begin{figure}[tp]
	\centering
		\includegraphics[width=0.48\textwidth]{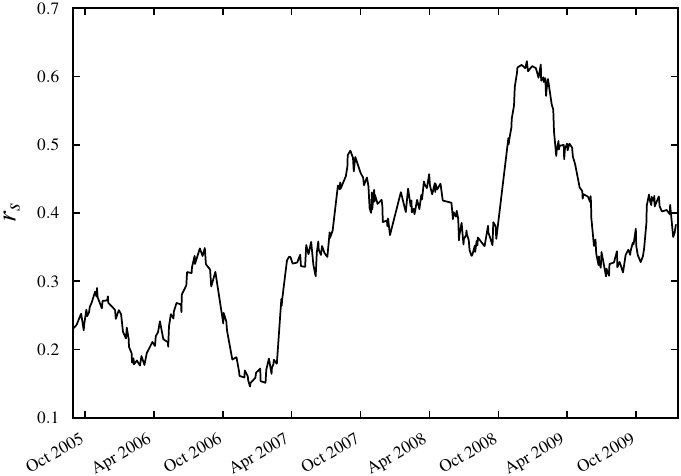}
	\caption{Mean of pairwise Spearman's correlation coefficients for the dataset from Sep 2008 to Nov 2009 each evaluated with a moving window of $50$ trading days. In Feb $2007$ the overall level of correlation increases. }
	\label{fig:meancor}
\end{figure}

Furthermore, we are able to identify a regime switch in the correlation structure at the end for Feb $2007$, indicated by a shift from light to dark shared areas. This transition is reflected in a raised average correlation level which affects the measure of similarity. Figure \ref{fig:meancor} shows the average correlation of the $471$ assets over time. Between Feb 2007 and Apr 2007 the overall level of correlation increases, indicating the new correlation regime. The sharp transition on Feb 2007 was induced by large overall price drop of the stocks in the S\&P 500. This originated in drastic events on the chinese stock market\footnote{See, e.g., Cover Story of Bloomberg Businessweek, Mar 12 2007: \emph{What The Market Is Telling Us}.}.

\section{Similarity Weighted Estimators}
\label{sec.weightcorr}
The similarity measure $\zeta^{L}$ may serve as a weighting scheme for estimators of correlation or covariance matrices.
With respect to the reference point $t_{0}$ the scheme inscribes high weights to periods where the market behaved in a similar manner. On the other hand, the periods in which the market behaved very differently are suppressed.
Therefore, consider the adapted similarity measure
\begin{eqnarray}
\tilde{\zeta}^{L}(t, t_0) = 1-\frac{\zeta^{L}(t, t_0)}{2(K-1)}\,\,\,\,\,\,\,, t\in[t_0-T,t_0],
\end{eqnarray}
where $T$ is the total number of considered time steps, i.e., the length of the time series.
The factor $K$ refers to the number of assets to include. It is easily checked that $2(K-1)$ represents the theoretical maximum possible value of $\zeta$, i.e., the highest possible dissimilarity.

We note that the probe matrices ${C}^{L}$ in equation (\ref{eq:simmeasure}) are estimated with window length $L.$ Therefore, within the timespan $[t_{0}-L,t_{0}],$ they share identical values with the probe matrix at $t=t_0.$  $\tilde{\zeta}^{L}(t, t_0)$ is then dominated by the amount of identical values and not by the estimated similarity.
Therefore, the similarity measure is not reliable within this region and is set to the maximum value of the other timespans, resulting in a corrected measure
\begin{eqnarray}
\tilde{\zeta}^{*L}(t,t_0)=
\begin{cases}
\max(\tilde{\zeta}^{L}(t<t_{0}-L,t_0)) & t \in [t_{0}-L,t_{0}] \\
\tilde{\zeta}^{L}(t,t_{0}) & t \in [t_0-T,t_{0}-L[\ .
\end{cases}
\end{eqnarray}
A normalized weighting scheme for the estimation of the correlation or covariance matrix $C(t_0)$ or $\Sigma(t_0)$ at time $t=t_0$ is then
\begin{eqnarray}
w(t,t_{0},L)= \tilde{\zeta}^{*L}(t,t_0) / \left(\sum\limits_{t=t_0-T}^{t_0} \tilde{\zeta}^{*L}(t,t_0) \right)\ ,
\end{eqnarray}
resulting in the weighted estimators
\begin{align}
\widehat{C}(t_{0}) &= \sum \limits_{t=t_0-T}^{t_{0}} w(t, t_{0},L)\ \widehat{C}^{L}(t)\quad \mathrm{and} \nonumber \\
\widehat{\Sigma}(t_{0}) &= \sum \limits_{t=t_0-T}^{t_{0}} w(t, t_{0},L)\ \widehat{\Sigma}^{L}(t)\ .
\label{eq:simweightedcorr}
\end{align}
The superscript $L$ again denotes the respective window length of the estimators.
For large $T$ and time series with dynamic correlation structure, the weighting scheme should be restricted to the $s$ largest values of $w.$ This leads to a complete suppression of dissimilar parts of the data. Let $w_{(s)}$ denote the $s$-th largest value of $w.$ The restricted scheme $w_s$ is then given by
\begin{eqnarray}
w_{s}(t, t_{0},L)=  |w-w_{(s)}|_+/\sum_{t=t_0-T}^t|w-w_{(s)}|_+\,
\label{eq:simweightedcorrmax}
\end{eqnarray}
with $|w-w_{(s)}|_+=\max(w(t, t_{0},L)-w_{(s)} \,, \ 0).$

The unbiasedness of the estimators (\ref{eq:simweightedcorr}) in time series, where the underlying correlation matrix is constant, is easily checked. However, due to fluctuations of the weights $w,$ their variances are expected to be slightly larger than for a constant non-adaptive weighting scheme $w=1/T.$ These effects are explored in the simulation study in the next section.

\section{Simulation Study}
\label{sec.simStud}

The study presented here aims at the validation of the estimators introduced in the last section. We estimate correlation and compare it to standard estimators with respect to the bias and variance. The study consists of $3$ scenarios of normally distributed daily returns of $16$ assets. The scenarios are constructed similarly to the testing environments in \cite{Pafka04}.

The first scenario is equicorrelation with equicorrelation parameter $\rho=0.7.$ This means that all pairwise correlations of the correlation matrix $(\rho_{i,j})$ of the $16$ asset returns are equal to $\rho=0.7$ for $i\neq j.$
\begin{figure}
	\centering
		\subfigure[Scenario 2]
		{
		\includegraphics[width=0.225\textwidth]{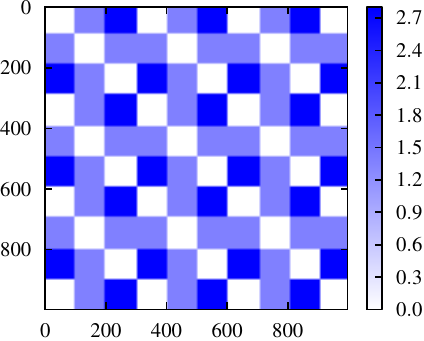}
		}
		\subfigure[Scenario 3]
		{
		\includegraphics[width=0.225\textwidth]{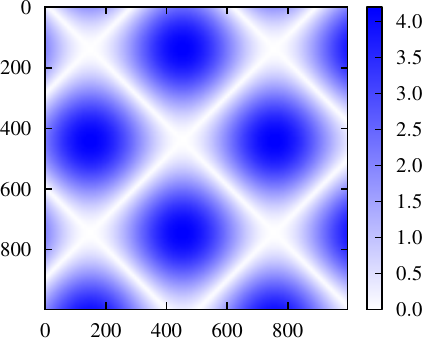}
		}
	\caption{Shown are the theoretical similarity matrices of the second and third scenario for the first $1000$ trading days. In the left figure, the discrete regimes of the second scenario are clearly visible. In the right figure the similarity matrix of the third scenario is shown. It shows no sudden changes.}
	\label{fig:simsim}
\end{figure}
In the second and third scenarios, the market consists of two equicorrelated branches, the first 8 assets with equicorrelation parameter $\rho_1$ and the second 8 assets with $\rho_3$. Assets of two different branches are equicorrelated with equicorrelation parameter $\rho_2=0.2.$
The equicorrelation parameters of the branches change over time, i.e., $\rho_1=\rho_1(t)$ and $\rho_3=\rho_3(t).$

In the second scenario, the market switches deterministically in turn between three different regimes.
Each regime lasts $100$ trading days. In regime $1$, the branches are equicorrelated with parameters $\rho_1=0.7$ and $\rho_3=0.3.$ In the second regime, these parameters are both equal to $0.5,$ in the third regime, they are $0.3$ and $0.7,$ respectively.

In the third scenario, the parameters $\rho_1(t)$ and $\rho_3(t)$ change sinusoidally with the trading days $t$ according to: \begin{align*}\rho_1(t)&=0.4+0.3\sin\left(\frac{t}{600}2\pi\right) \\ \rho_2(t)&=0.4+0.3\sin\left(\frac{t-300}{600}2\pi\right).\end{align*}
Figure \ref{fig:simsim} depicts the theoretical similarity matrices of the second and third scenario for the first $1000$ trading days. The discrete regimes of the second scenario are clearly visible while the similarity matrix of the third scenario shows no sudden changes.

The similarity weighted estimator is compared to benchmark estimators. The first benchmark is the standard Pearson correlation estimator based on the last $300$ returns. As the second benchmark, we use the RiskMetrics exponentially weighted correlation estimator. The estimator weights the $j$-th recent return with weight $w_j.$ The weights are chosen according to \begin{align*}w_j=\left(\frac{1-\lambda^n}{1-\lambda} \right)^{-1}\sum_{j=1}^n \lambda^{j-1} \ ,
\end{align*} as suggested by \cite{RiskMetrics}.

We estimate the correlation matrix in all three scenarios for the days $t=1000$, $t=2500$ and $t=5000.$  To estimate mean and variance of the estimators, each simulation is independently repeated $400$ times. The results are presented in tables \ref{tab:scen1} to \ref{tab:scen3} which show the means and sample standard deviations of the parameters of interest over the $400$ repetitions for the 3 estimators.

Table \ref{tab:scen1} shows the results of scenario $1$. The parameter $\rho_{i,j}=0.7$ is estimated adequately in all cases, which confirms that all estimators are unbiased in this setup. As to be expected, the standard estimator has the lowest variance. It uses a constant weighting scheme. This is known to be optimal, when the underlying correlation is constant. In this setup, the adaptive weighting scheme of the similarity weighted estimator should also equally weight all observations. However, due to stochastic fluctuations the weights vary. Therefore, the variance of the estimator is slightly larger than the variance of the standard estimator. The exponential weighted scheme suffers the highest variance as it heavily weights the most recent observations. This results in an unbalanced weighting scheme which is not optimal in this scenario.

The results of scenario $2$ are shown in table \ref{tab:scen2}. The standard estimator is highly biased since its weighting scheme weights data from all $3$ regimes equally. Unlike the standard estimator, the exponential estimator weights the most recent observations most and therefore seems unbiased. Again, its variance is the largest among the three considered estimators. The similarity weighted estimator shows variances comparable to the variance of the unweighted estimator but is nearly unbiased. Table \ref{tab:scen3} shows the results of scenario $3.$ Since in this scenario the true parameters change continuously in time, the scenario tests if the adaptive scheme given by the similarity measure separates the similar regions from the dissimilar ones in an adequate way. The results are analogue to the results of scenario $2,$ but for the days $1000$ and $2500$ also the similarity weighted and exponentially weighted estimators deviate from the theoretical values. However, they both are much closer to the theoretical value than the unweighted estimator. 

It is worthwhile to note that in all scenarios the bias of the similarity weighted estimator is similar to the bias of the exponentially weighted estimator. The standard deviation of the similarity weighted estimator, however, is only slightly larger than the standard deviation of the unweighted estimator and much smaller than the standard deviation of the exponentially weighted estimator.

\begin{table}

\begin{tabular}{llccccccccc}
\toprule
&  &  & \multicolumn{2}{c}{similarity} & & \multicolumn{2}{c}{unweighted} &
& \multicolumn{2}{c}{exponential} \\ \cline{4-5}\cline{7-8}\cline{10-11}
day& ${\rho}$ &  & $\widehat{{\rho}}$ & $\widehat{{\mathbf{\sigma}}}_{\widehat{{\rho }}}$
&  & $\widehat{{\rho}}$ & $\widehat{{\mathbf{\sigma}}}_{\widehat{{\rho }}}$ &  & $\widehat{{\rho}}$ & $\widehat{{\mathbf{\sigma}}}_{\widehat{{\rho }}}$ \\
\colrule
1000 & 0.7 &  & 0.6974 & 0.0364 &  & 0.6979 & 0.0296 &  & 0.6947 & 0.0736 \\
2500 & 0.7 &  & 0.6991 & 0.0403 &  & 0.7002 & 0.0288 &  & 0.6977 & 0.0729 \\
5000 & 0.7 &  & 0.6973 & 0.0429 &  & 0.7004 & 0.0296 &  & 0.7022 & 0.0718\\
\botrule
\end{tabular}
\caption{Simulation results for scenario 1, the scenario of constant correlation structure. Shown are the results for the similarity weighted, the unweighted and the exponentially weighted estimator. All estimators are unbiased, the exponentially weighted estimator shows the largest standard deviation.}
\label{tab:scen1}
\end{table}
\begin{table}
\begin{tabular}{llccccccccc}
\toprule
&  &  & \multicolumn{2}{c}{similarity} & & \multicolumn{2}{c}{unweighted} &
& \multicolumn{2}{c}{exponential} \\ \cline{4-5}\cline{7-8}\cline{10-11}
day & ${\rho}$ &  & $\widehat{{\rho}}$ & $\widehat{{\mathbf{\sigma}}}_{\widehat{{\rho }}}$
&  & $\widehat{{\rho}}$ & $\widehat{{\mathbf{\sigma}}}_{\widehat{{\rho }}}$ &  & $\widehat{{\rho}}$ & $\widehat{{\mathbf{\sigma}}}_{\widehat{{\rho }}}$ \\ 
\colrule
\multirow{3}{*}{1000}
&0.7 && 0.6605 & 0.0339 && 0.4992 & 0.0448 && 0.6911 & 0.0759\\
&0.2 && 0.2007 & 0.0528 && 0.1995 & 0.0552 && 0.1925 & 0.1350\\
&0.3 && 0.3368 & 0.0498 && 0.5000 & 0.0442 && 0.3002 & 0.1299\\
\colrule
\multirow{3}{*}{2500}
&0.7 && 0.6792 & 0.0341 && 0.4985 & 0.0456 && 0.6883 & 0.0734\\
&0.2 && 0.2026 & 0.0551 && 0.1996 & 0.0553 && 0.1975 & 0.1363\\
&0.3 && 0.3199 & 0.0464 && 0.4989 & 0.0443 && 0.3019 & 0.1288\\
\colrule
\multirow{3}{*}{5000}
&0.5 && 0.4992 & 0.0492 && 0.4972 & 0.0448 && 0.5010 & 0.1072\\
&0.2 && 0.2005 & 0.0623 && 0.1987 & 0.0558 && 0.1947 & 0.1384\\
&0.5 && 0.4994 & 0.0502 && 0.4995 & 0.0458 && 0.4946 & 0.1076\\
\botrule
\end{tabular}
\caption{Simulation results for scenario 2, the scenario of discrete regimes in the correlation structure. Shown are the results for the similarity weighted, the unweighted and the exponentially weighted estimator. Clearly, the similarity weighted and the exponentially weighted estimators are less biased than the unweighted estimator. The exponentially weighted estimator shows a much larger standard deviation than the similarity weighted one. Note that the theoretical values refer to the values of the regimes of one day before the mentioned days.}
\label{tab:scen2}
\end{table}

\begin{table}
\begin{tabular}{llccccccccc}
\toprule
&  &  & \multicolumn{2}{c}{similarity} & & \multicolumn{2}{c}{unweighted} &
& \multicolumn{2}{c}{exponential} \\ \cline{4-5}\cline{7-8}\cline{10-11}
day & ${\rho}$ &  & $\widehat{{\rho}}$ & $\widehat{{\mathbf{\sigma}}}_{\widehat{{\rho }}}$
&  & $\widehat{{\rho}}$ & $\widehat{{\mathbf{\sigma}}}_{\widehat{{\rho }}}$ &  & $\widehat{{\rho}}$ & $\widehat{{\mathbf{\sigma}}}_{\widehat{{\rho }}}$ \\ 
\colrule
\multirow{3}{*}{1000}
&0.1402 && 0.2144 & 0.0548 && 0.4941 & 0.0457 && 0.1927 & 0.1391\\
&0.2 && 0.1994 & 0.0524 && 0.1997 & 0.0561 && 0.1951 & 0.1374\\
&0.6598 && 0.5796 & 0.0429 && 0.3058 & 0.0532 && 0.5994 & 0.0869\\
\colrule
\multirow{3}{*}{2500}
&0.6598 && 0.6012 & 0.0478 && 0.3062 & 0.0540 && 0.6029 & 0.0891\\
&0.2 && 0.2007 & 0.0529 && 0.2007 & 0.0559 && 0.1992 & 0.1363\\
&0.1402 && 0.1994 & 0.0516 && 0.4935 & 0.0459 && 0.1905 & 0.1372\\
\colrule
\multirow{3}{*}{5000}
&0.6598 && 0.6692 & 0.0361 && 0.4955 & 0.0457 && 0.6767 & 0.0804\\
&0.2 && 0.1993 & 0.0553 && 0.1996 & 0.0552 && 0.1979 & 0.1360\\
&0.1402 && 0.1302 & 0.0469 && 0.3030 & 0.0539 && 0.1221 & 0.1397\\
\botrule
\end{tabular}
\caption{Simulation results for scenario 3, the scenario with sinusoidally changing correlation structure. Shown are the results for the similarity weighted, the unweighted and the exponentially weighted estimator. Clearly, the similarity weighted and the exponentially weighted estimators are less biased than the unweighted estimator. Due to the fast changing structures, for the days $1000$ and $2500$ they deviate from the theoretical values but are much closer to the theoretical value than the unweighted estimator. Again, the exponentially weighted estimator shows a much larger standard deviation than the similarity weighted one.}
\label{tab:scen3}
\end{table}

\section{Application to financial data}
\label{sec.emp}

In this section, we apply our estimator to financial data in the context of mean-variance portfolio allocation. The application is motivated by \cite{engle06} who showed that the realized volatility of theoretically optimal portfolios is lowest if the covariance matrices for the optimization process are correctly specified.
We therefore compare realized volatility and return of various portfolios drawn from the S\&P 500. The study shows that portfolios based on the similarity weighted estimator as discussed in this paper outperform alternative portfolios. We conclude that these similarity weighted estimators perform very well in real data applications.

The value $V$ of a portfolio consisting of $K$ assets with prices $S_{i}$ and corresponding portfolio weights $w_{i}$ ($i=1\dots K$) is given by
\begin{equation}
V=\sum\limits_{k=1}^{K}{w_k S_k} = \vec{w}'\vec{S}\ ,
\end{equation}
where $\vec{S}$ refers to the $(K \times 1)$ vector of asset prices and $\vec{w}$ contains the respective weights.

Consider an investment period from day $t=0$ to day $t=T.$
Let $\mathbf{\Sigma}$ and $\vec{\mu}$ be covariance matrix and the expectation of the $K$ asset returns over the period. Then portfolio variance and expectation at time $t=T$ are given by
\begin{align*}
\text{Var}(V_T)&= \vec{w}'\mathbf{\Sigma}\vec{w}\ , \\
E[V_T]&=V_0(\vec{1}+\vec{w}'\vec{\mu})\ ,
\end{align*}
where ${\bm{1}}$ is a vector of ones. Let $\delta V_t$ denote the daily returns of the portfolio over the investment period. Then \begin{align*}RV=\sum_{t=0}^T (\delta V_t)^2\end{align*} is the realized volatility of the portfolio which is a measure of the portfolios' risk over the investment period.

In mean-variance portfolio optimization as introduced by \cite{markowitz52}, optimal portfolio weights $w_i$ are derived by minimization problems of the form
\begin{equation}
\label{glg.markow}
\underset{\vec{w}}{\mathrm{min}}\left\{\frac{1}{2}\vec{w}'\mathbf{\Sigma}\vec{w} - \gamma\vec{w}'\vec{\mu} \   \right\}\ ,
\end{equation}
subject to certain constraints, e.g.,
\begin{equation}
\sum\limits_{k=1}^{K}w_{k}=1
\end{equation}  (budget restriction).
 The parameter $\gamma>0$ is the investors' \emph{risk tolerance parameter}. A value $\gamma=0$ denotes no risk tolerance. In this case, the investor's only aim is to minimize the portfolio variance. Large values of $\gamma$ denote risk neutrality, i.e., the investor maximizes the expected portfolio return only.

Since different investors have different risk tolerance levels, we focus on two special cases of the minimization problem. We consider the minimum-variance portfolio (\emph{MVP}), i.e., the portfolio of minimal variance without further constraints, and the portfolio with minimal variance under the constraint of a fixed target portfolio return $R$ (\emph{TRP}).
The minimum-variance portfolio is the solution of equation (\ref{glg.markow}) when $\gamma$ is set to zero, i.e., the investor is not risk tolerant. To obtain portfolio $TRP,$ $\gamma$ can easily be expressed by the target return $R$
\begin{equation}
\gamma=\frac{ R-\frac{\alpha}{\beta}}{\vec{\mu}' \mathbf{\Sigma}^{-1}\vec{\mu}-\frac{\alpha^{2}}{\beta}}\ ,
\end{equation}
where $\alpha={\mathbf{1}}'\mathbf{\Sigma}^{-1}\vec{\mu}$ and
$\beta={\mathbf{1}}'\mathbf{\Sigma}^{-1}{\mathbf{1}}$.

In a recent paper, \cite{kritzman10} argue that minimum-variance portfolios outperform various other strategies of portfolio optimization, even with respect to the their return.

By contrast, \cite{demiguel09} raise the question whether portfolio optimization pays out at all. In their results, optimized portfolios do not significantly outperform naive diversified portfolios, i.e., portfolios where the same amount $1/n$ is invested in $n$ assets.
We therefore include this naive portfolio in our study, even though the naive portfolio does not depend on estimators of correlation or covariance. The portfolio strategies \emph{MVP} and \emph{TRP} allow to rank the estimators of the covariance structure according to the portfolio performance, while the outcomes of the naive portfolio confirm the overall plausibility of the results.

The basic idea of the study is to  calculate optimal portfolios for every day of our dataset and to evaluate them over some investment horizon $T$ with respect to risk (realized volatility) and return. We then compare the results of the different strategies and estimators.

\begin{figure}[tbp]
	\centering
		\subfigure[14 day evaluation]
		{
		\includegraphics[width=0.48\textwidth]{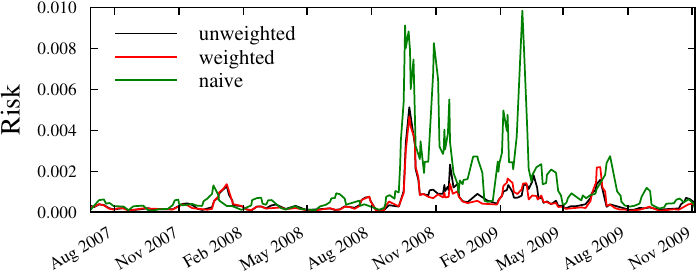}
		}\\
		\subfigure[28 day evaluation]
		{
		\includegraphics[width=0.48\textwidth]{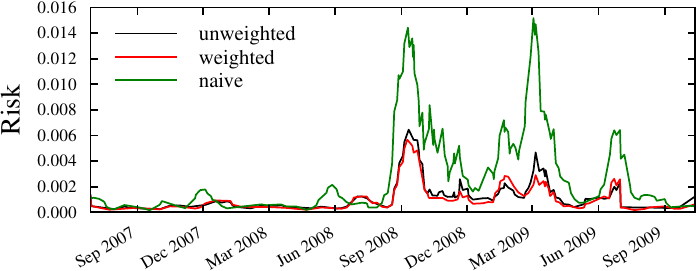}
		}\\
		\subfigure[56 day evaluation]
		{
		\includegraphics[width=0.48\textwidth]{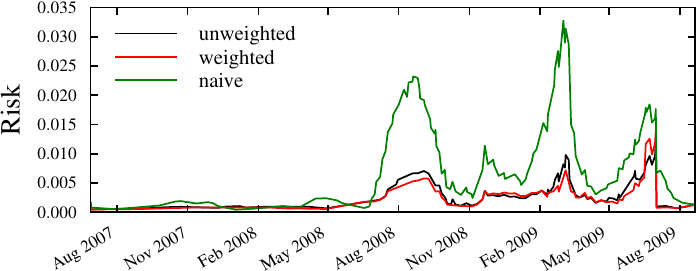}
		}
	\caption{Average realized volatility risk for a minimum variance portfolio over $10$ portfolio constellations. The unweighted correlation matrix is compared to a weighted correlation matrix using the similarity measure $\zeta^{50}$. The results are compared to a naive portfolio as a reference.}
	\label{fig:realizedportfolios2}
\end{figure}

\begin{table}[bt]
\begin{tabular}{rrrr}
\toprule
  & \multicolumn{3}{c}{Optimization type} \\
\colrule
 Evaluation  & unweighted & weighted & naive\\
\colrule
14 day  & 0.00052 & 0.00047 & 0.00124 \\
\colrule
28 day & 0.00102 & 0.00090 & 0.00252 \\
\colrule
56 day  & 0.00231 & 0.00211 & 0.00593 \\
\botrule
&&&
\end{tabular}
\caption{Average realized risk in mean-variance portfolio optimization for the minimum variance portfolio and different evaluation windows. The last column provides a comparison to the naive portfolio.}
\label{tab:minvar}
\end{table}
We use the same dataset as in section \ref{sec.similarity}, i.e., the $471$ assets of the S\&P $500$ index that are included in the index from $2005$ to the beginning of $2010$.
From this dataset, we randomly choose 10 portfolio constellations of $100$ stocks each. For every trading day from Aug $2008$ to Nov $2009$ we compute portfolio weights for the constellations regarding to the $3$ strategies. The necessary covariance estimates rely on the similarity weighted estimator and alternatively on the unweighted estimator. For the first estimator, we need a similarity measure, which is determined as discussed in section \ref{sec.similarity}. The probe matrices to calculate the similarity measure rely on moving windows of $L=50$ trading days and are based on all $471$ assets of the dataset. Window lengths between $30$ to $70$ trading days lead to similar results. However, the results for window lengths around $50$ seem to be quite representative. The weighting scheme of the estimator includes the $s=300$ most similar past days. The unweighted estimator is based on a moving window of $300$ days. The weights of the target return portfolio rely on an additionally specified target return $R$ and on estimates of the vector $\vec{\mu}$ of expected returns as well. The vector $\vec{\mu}$ is estimated by the returns of the portfolio's stocks for every trading day from a moving window of $14$ trading days.

The target return is then adaptively chosen to be $5$ percentage points above the average entree of $\vec{\mu}.$

\begin{figure}[tbp]
	\centering
		\subfigure[14 day evaluation]
		{
		\includegraphics[width=0.48\textwidth]{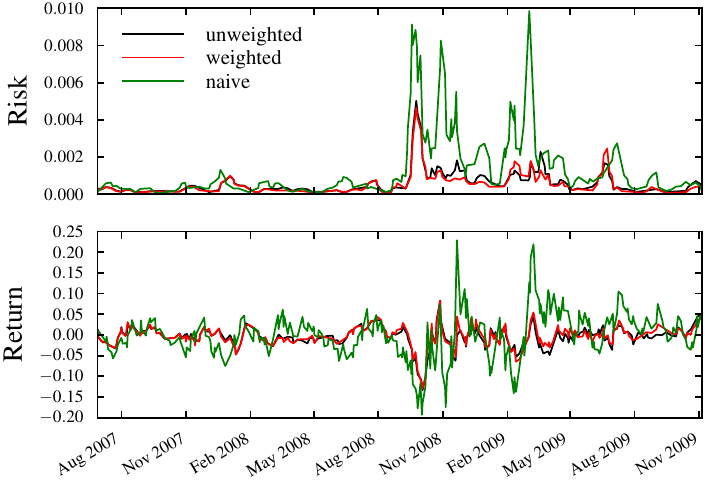}
		}\\
		\subfigure[28 day evaluation]
		{
		\includegraphics[width=0.48\textwidth]{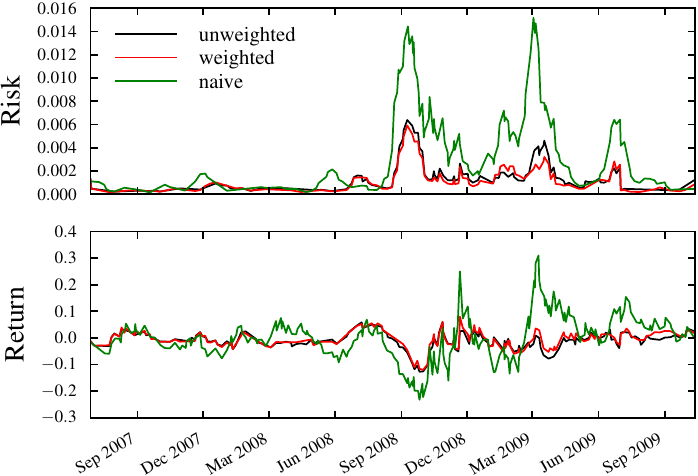}
		}\\
		\subfigure[56 day evaluation]
		{
		\includegraphics[width=0.48\textwidth]{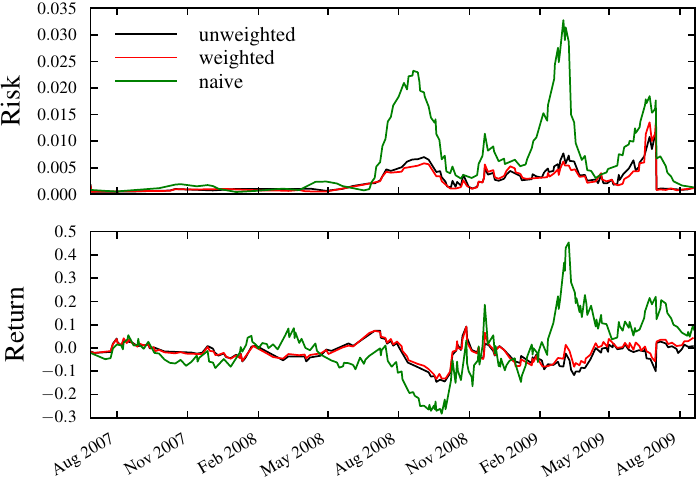}
		}
	\caption{Average realized return and realized volatility (risk) for a target return of 5\% above portfolio drift over $10$ portfolio constellations. The unweighted correlation matrix ($300$ days moving window) is compared to a similarity weighted correlation matrix using the similarity measure $\zeta^{50}$. The results are compared to a naive portfolio as a reference.}
	\label{fig:realizedportfolios}
\end{figure}

The evaluation results of realized volatility and returns are shown in Fig. \ref{fig:realizedportfolios}, \ref{fig:realizedportfolios2} and tables \ref{tab:minvar} and \ref{tab:realized}.
The evaluation periods are $14$, $28$ and $56$ trading days, respectively. The results shown are averages of the $10$ portfolio constellations. Visual inspection of the figures shows that the naive portfolio performs worst, especially during the financial crisis. In that time, the incorporation of the covariance structure into the portfolio weights pays out. Realized volatility of the optimized portfolios consistently lies below the realized volatility of the naive portfolios whereas the similarity weighted scheme obtains the best results. The results are robust for the considered investment horizons which is shown in the tables in more detail.

In both cases, in the minimum variance portfolio (\emph{MVP}) as well as in the 5\% above market drift portfolio (\emph{TRP}), the similarity weighting significantly reduces the realized risk. Moreover, the \emph{TRP} case reveals that the realized return could be improved compared to the unweighted optimization, although the naive portfolio features an even higher return.

\begin{table}[bt]
\begin{tabular}{rrrrr}
\toprule
 & & \multicolumn{3}{c}{Optimization type} \\
\colrule
 Evaluation & & unweighted & weighted & naive\\
\colrule
14 day & Risk & 0.00054 & 0.00048 & 0.00124 \\
& Return & -0.00391 & -0.00252 & -0.00103 \\
\colrule
28 day & Risk & 0.00108 & 0.00095 & 0.00252 \\
& Return & -0.00918 & -0.00567 & -0.00202 \\
\colrule
56 day & Risk & 0.00244 & 0.00224 & 0.00593 \\
& Return & -0.02138 & -0.01547 & -0.00362 \\
\botrule
&&&
\end{tabular}
\caption{Average realized return and realized risk in mean-variance portfolio optimization for a target return of 5\% above the market drift and different evaluation windows. The last column provides a comparison to the naive portfolio.}
\label{tab:realized}
\end{table}

\section{Conclusion}
\label{sec.concl}
We introduced a measure that quantifies the similarity of the correlation structure for two different times. This measure gives a clear indication for drastic changes in the market structure as past the beginning of the financial crisis 2008.

This measure was adapted to calculate weighted correlation and covariance matrices in which information that originated from a similar market state is weighted higher.

We analyzed the resulting similarity weighted estimators in a simulation study and applied it to a mean-variance portfolio optimization in a historical study. The results show that our method reduces the portfolio volatility as well as it enhances the realized return compared to the use of unweighted correlations. The application of similarity weighted estimators is especially advantageous in periods in which the market structure changes drastically.

\section*{Acknowledgement}
M.C.M. acknowledges financial support from Studienstiftung des deutschen Volkes.

\section*{References}
\bibliographystyle{rQUF}
\bibliography{arxiv.bbl}
\end{document}